\pdfoutput=1

\documentclass[11pt]{article}

\usepackage[]{emnlp2021}

\usepackage{times}
\usepackage{latexsym}
\usepackage{graphicx}
\usepackage{relsize}

\usepackage[T1]{fontenc}

\usepackage[utf8]{inputenc}

\usepackage{microtype}

\title{Changes in Twitter geolocations: Insights and suggestions for future usage}

\author{Anna Kruspe \\
  \smaller{Technical University of Munich} \\
  \texttt{\smaller{anna.kruspe@tum.de}} \\\And
  Matthias Häberle \\
  \smaller{German Aerospace Center (DLR)} \\
  \texttt{\smaller{matthias.haeberle@dlr.de}} \\\And
  Eike J. Hoffmann \\
  \smaller{Technical University of Munich} \\
  \texttt{\smaller{eike.jens.hoffmann@tum.de}} \\\AND
  Samyo Rode-Hasinger \\
  \smaller{Technical University of Munich} \\
  \texttt{\smaller{samyo.rode@tum.de}} \\\And
  Karam Abdulahhad \\
  \smaller{German Aerospace Center (DLR)} \\
  \texttt{\smaller{karam.abdulahhad@dlr.de}} \\\And
  Xiao Xiang Zhu \\
  \smaller{German Aerospace Center (DLR)} \\
  \texttt{\smaller{xiaoxiang.zhu@dlr.de}} \\
  }

\begin{document}
\maketitle
\begin{abstract}
Twitter data has become established as a valuable source of data for various application scenarios in the past years. For many such applications, it is necessary to know where Twitter posts (tweets) were sent from or what location they refer to. Researchers have frequently used exact coordinates provided in a small percentage of tweets, but Twitter removed the option to share these coordinates in mid-2019. Moreover, there is reason to suspect that a large share of the provided coordinates did not correspond to GPS coordinates of the user even before that.

In this paper, we explain the situation and the 2019 policy change and shed light on the various options of still obtaining location information from tweets. We provide usage statistics including changes over time, and analyze what the removal of exact coordinates means for various common research tasks performed with Twitter data. Finally, we make suggestions for future research requiring geolocated tweets.
\end{abstract}

\section{Introduction}
Twitter data has become an invaluable source of information for a range of application scenarios. In many situations, messages posted on this platform can provide insights faster and on a more fine-grained level than any other source of information. Moreover, it is a gigantic source of opportunistic, and therefore cheap, data. Example applications include situational awareness in disaster situations, where Twitter users provide updates much faster than official news sources or satellite imagery \cite{Kruspe2021}; collection of personal opinions and insights into human behavior, where Twitter is faster, more expansive, and cheaper than traditional surveys \cite{Ceron2014}; or mapping of human settlements, where Twitter users can provide information that is difficult or impossible to obtain from any other source \cite{Haeberle2019}.

One crucial factor when analyzing Twitter posts is the ability to align these messages to places around the world. In disaster situations, knowing what location a certain tweet refers to can be a matter of life or death \cite{Singh2019}. In mapping tasks, information gained from Twitter is only valuable when it can be placed on the requested level of detail, e.g. buildings \cite{Terroso_Saenz_Munoz_2020}. On a broader scale, even knowing what city or country a tweet was posted from can provide significant insights into regional differences and developments, such as in the ongoing COVID-19 situation \cite{Kruspe_Haeberle_Kuhn_Zhu_2020}.

For these reasons, a lot of social media-focused research relies on geolocations provided within Twitter data. Up until mid-2019, precise geolocations were reliably available for a sufficient subset of tweets. On June 18th, 2019, however, Twitter announced they would remove the option to attach precise geolocations to tweets (geotagging)\footnote{\url{https://twitter.com/TwitterSupport/status/1141039841993355264}}\cite{Hu2020}. The reasoning given at the time was that not many users were taking advantage of this feature. Privacy concerns in connection with precise geolocations have also been voiced in the past \cite{Park_Seglem_Lin_Zufle_2017,Fiesler_Proferes_2018}, so there is a possibility that these issues also factored into the policy change. In this paper, we will discuss how the situation has changed and what this means for Twitter-based research going forward.

Besides analyzing the current geolocation situation in Twitter data, we aim to shed light on the amount of non-used tweets simply because they do not have point coordinates. Although Twitter data may not the best choice for fine-grained location based research, as we will see, Twitter still represents a treasure trove of geolocated information. Whatever we attempt to do with it, we do need to take into account the required and available location granularity. With this paper, we hope to pave the way for more solid and realistic Twitter based research. Our main contributions are (1) a detailed description of the change and current status, (2) statistics on the availability of different kinds of geolocations, and (3) a detailed reflection on the consequences for various research and how to deal with them.

In the next section, we first discuss in detail what exactly Twitter's policy change entailed, and present experimental results to determine more closely which geotagging options are currently available and what the resulting data looks like. Section \ref{sec:stats} provides statistics on the availability of various types of geolocations since 2019. In sections \ref{sec:effect} and \ref{sec:suggestions}, we analyze what effect the changes have on research, and make suggestions for adapting future research tasks accordingly. Section \ref{sec:conclusion} provides a conclusion.

\section{Geolocation availability}\label{sec:availability}
In the following, we will use the term ``geolocated'' to mean tweets containing explicit metadata about a geographic location they were posted from or are referring to, and ``geotagging'' as the user action that causes this metadata to be attached.
\subsection{Data format and situation before mid-2019}
In research, Twitter data is commonly obtained via the Twitter Streaming API in JSON format. In this format, each tweet is represented via a fixed set of attributes containing all of its public information. This format essentially contains two attributes for geotagging\footnote{\url{https://developer.twitter.com/en/docs/tutorials/filtering-tweets-by-location}}\footnote{\url{https://developer.twitter.com/en/docs/twitter-api/v1/data-dictionary/object-model/geo}}: 
\begin{description}
\item[\texttt{coordinates}] containing the fields \texttt{type}, which will always be \texttt{Point}, and \texttt{coordinates}, which contains longitude and latitude values.
\item[\texttt{place}] containing the fields \texttt{id, url, place\_type, name, full\_name, country\_code, country}, and \texttt{bounding\_box}. \texttt{name, full\_name, country\_code}, and \texttt{country} provide human-readable semantic information about a place, while \texttt{place\_type} can be ``country'', ``city'', ``poi'', etc. \texttt{bounding\_box} contains a set of coordinates spanning a polygon, which may have surface 0, i.e. be a point.
\end{description}
For completeness, we also want to mention that an attribute called \texttt{geo} exists, but it is now deprecated.

By its original definition, \texttt{coordinates} is supposed to contain the exact geolocation where a tweet was posted. Before mid-2019, this meant that when a user gave Twitter permissions to use their location for geotagging, this attribute was filled with the coordinates obtained from the device that the user was posting from, particularly its GPS module. If those permissions were not given, the attribute was simply set to \texttt{null}.

In contrast, the \texttt{place} attribute serves to assign a pre-defined geographic entity to a post. Twitter offers users the option to select this entity from a list of those found nearby (within a radius of roughly 200m). These entities may be countries, cities, neighborhoods, points of interest, etc. \texttt{place}'s sub-fields are then automatically filled using information from geolocation services provided by Foursquare or Yelp\footnote{\url{https://help.twitter.com/en/using-twitter/tweet-location}}.

User profiles can also contain geolocations; this is the case for around 30-40\% of profiles\footnote{\url{https://developer.twitter.com/en/docs/tutorials/advanced-filtering-for-geo-data}}. This field can be freely set by the user, and may therefore contain fictitious or nonsensical values. Automatic geocoding is performed for plausible values, leading to information similar to the \texttt{place} attribute described above within the \texttt{user.derived.locations} field. Unfortunately, this information can only be obtained via the paid Enterprise API\footnote{\url{https://developer.twitter.com/en/docs/twitter-api/enterprise/enrichments/overview/profile-geo}}. For a detailed look at geotagging behavior of users pre-2019, see \cite{Huang2019, Tasse2017}.

\subsection{Situation since mid-2019}
The policy change in mid-2019 did not affect the \texttt{place} attribute; the option to set this still exists. However, the \texttt{coordinates} attribute now cannot be filled anymore, at least not when using Twitter's own clients for posting.

Twitter's original announcement stated that users will ``still be able to tag your precise location in Tweets through our updated camera''. However, in our own experiments (conducted on iOS with the latest German Twitter app), we were not able to confirm this option.

\subsection{Cross-posts from third-party sources}
Besides its direct clients, Twitter also provides options for cross-posting from various other social media platforms. Table \ref{tab:sources} provides an overview of the distribution of the sources of tweets collected in the time-frame of August 3\textsuperscript{rd} to August 9\textsuperscript{th}, 2021. The total share of tweets posted from Twitter's own clients is around 95\%.
\begin{table}[h!]
    \centering
    \begin{tabular}{l r c }
        \hline
         & \#tweets & \%total \\
        \hline
        \hline
        Total & 25,756,667 & 100\% \\
        \hline
        Twitter for Android & 11,657,427 & 45\% \\
        Twitter for iPhone & 9,410,305 & 37\% \\
        Twitter Web Client & 2,982,670 & 12\% \\
        Twitter for iPad & 506,638 & 2\% \\
        Twitter for Mac & 12,040 & 0.05\%\\
        \hline
        Instagram & 22,151 & 0.09\%\\
        Foursquare (+ Swarm) & 2,034 & $\ll$0.01\%\\
        CareerArc 2.0 & 1,229 & $\ll$0.01\%\\
        Others & 1,162,173 & 4.5\%\\
        \hline
    \end{tabular}
    \caption{Distribution of tweet sources. ``Others'' mainly includes Twitter bots}
    \label{tab:sources}
\end{table}

\paragraph{Instagram} As the statistic reveals, Instagram is by far the most common source of cross-posts on Twitter. We therefore analyzed the options for geotagging in Instagram and their result in the Twitter data format. Just like the native Twitter apps, Instagram allows users to pick a geolocation for their posts at various granularities (e.g. point of interest (POI), city, etc.). No clear statement of the source of these locations has been made available by the company, but it seems likely that they are provided by the "Places Graph" service of Instagram's parent company Facebook. In contrast to Twitter's approach, Instagram users are allowed to pick locations from anywhere around the world.

When a user chooses a location on Instagram for a Twitter cross-post, both the \texttt{place} and the \texttt{coordinates} attributes are filled. The \texttt{place} attribute is always set to the city of the post and completed accordingly. The \texttt{coordinates} attribute will contain a single point coordinate, which is forwarded from Instagram. In our experiments, these were coordinates within the location picked on Instagram. Consequently, the \texttt{coordinates} attribute now fulfills a different role than it originally did on Twitter; it is not anymore representative of the user's geolocation from which the post was sent, but of some pre-defined location selected by the user, which may be very different from their physical location.

\paragraph{Others}
The second-most frequent (by a large margin) external source, Foursquare, also allows its users to cross-post to Twitter, e.g. via its ``Swarm'' app. This was the only option that actually allowed us to create Twitter posts with an exact geolocation. However, this option required turning off several system-side privacy settings, and was difficult to use. We therefore do not expect many users to do this. Alternatively, the app allows users to select an arbitrary coordinate for their posts, which is the more likely provenance of geolocations in Foursquare-based tweets.

The third-most frequent third-party source of tweets is Career Arc 2.0, a social media recruiting service. This service is only available to business partners, and only within the USA. We were therefore not able to directly test how selected geolocations are mapped to the Twitter format, but posts in our sample were generally on the city level.
Due to the limited usage of this source in the dimensions of geography, use case, and user base, we do not see these tweets as a generally valuable source of information.

\subsection{Ethical Considerations / Motivations}
While the use of precise coordinates might be relevant to specifically orientated research, the General Data Privacy Regulation \cite{EC2019} highlights the importance of data minimization when mining personal data. Personal data is defined as any information relating to an identified or identifiable living individual \cite{EC2018}. Thus, personal data does include the geolocalization of a person. Data minimization translates into the reduced collection of personal data to the absolute minimum needed amount and variety to answer a specific research question. The removal of precise geolocations by Twitter falls in line with this goal, and researchers must keep these considerations in mind when collecting Twitter data for their specific purposes. This is particularly critical when dealing with tasks that attempt to analyze or make predictions on a user basis.

\section{Statistics}\label{sec:stats}
In this section, we perform statistical analyses to gather insights into what geolocation information is currently contained in tweets, and how the policy change impacted the availability of this information.
\paragraph{Share of geolocated tweets over-all}
\begin{figure}[!htb]
    \centering
    \includegraphics[width=0.48\textwidth]{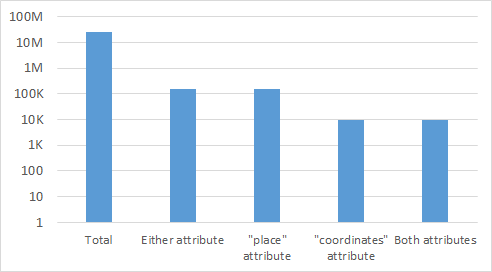}
    \caption{Shares of tweets with geolocations in one-week sample from August 2021, note log scaling (``Total'' = all tweets collected in the 1\% sample).}
    \label{fig:geoloc_availability}
\end{figure}
We first calculated the shares out of all tweets that contain any sort of geolocation on a sample collected from the free 1\% worldwide Twitter stream between August 3\textsuperscript{rd} and August 9\textsuperscript{th}, 2021. The results are shown in figure \ref{fig:geoloc_availability}. About .06\% of all tweets are geolocated. Nearly all of these have a filled \texttt{place} attribute, while only 6\% of geolocated tweets provide the \texttt{coordinates} attribute. The last two bars are nearly identical, i.e. if the \texttt{coordinates} attribute is filled, the \texttt{place} attribute is almost certainly also filled.


\paragraph{\texttt{coordinates} attribute}
\begin{figure}[!htb]
    \centering
    \includegraphics[width=0.48\textwidth]{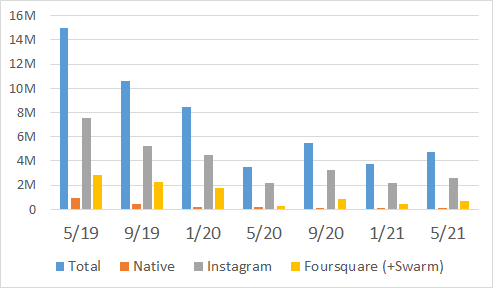}
    \caption{Sources of tweets with the \texttt{coordinates} attribute set by month (``Total'' = all tweets with \texttt{coordinates} collected in the 1\% sample of that month).}
    \label{fig:coordinates_sources}
\end{figure}
\begin{figure}[!htb]
    \centering
    \includegraphics[width=0.48\textwidth]{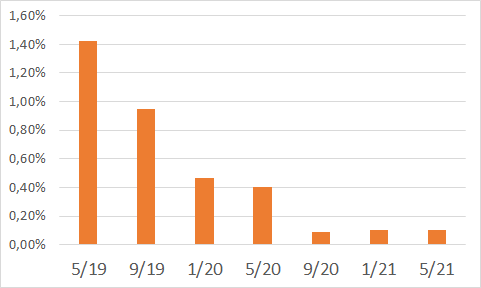}
    \caption{Percentages of geolocated tweets from native Twitter sources with the \texttt{coordinates} attribute set by month.}
    \label{fig:coordinates_native}
\end{figure}
Next, we took a closer look at the \texttt{coordinates} attribute. Figure \ref{fig:coordinates_sources} shows the sources of tweets that provide this attribute for every third month between May '19 and May '21, while figure \ref{fig:coordinates_native} shows the percentage of geolocated tweets coming from native Twitter clients where \texttt{coordinates} is set. As expected, the numbers for tweets from Twitter's own clients have been decreasing over time (we believe the reason that they did not drop immediately may have been due to usage of outdated versions). Even before the policy change, most tweets with a \texttt{coordinate} attribute came from Instagram. As explained above, this means that the assumption that this attribute provides users' exact geolocations was never correct for a large percentage of them. Surprisingly, we also see a drop in geolocation provision from other apps. There are several factors at play here. First of all, a seasonal fluctuation is normal due to vacation seasons \cite{Maurer2020}. Second, from May '20 onward, the COVID-19 pandemic most likely changed users' posting behavior with regards to their location. Finally, newer versions of mobile operating systems put their users’ privacy more into focus. Both iOS and Android made access to location services more visible and transparent with opt-ins for data sharing and notifications when location services were requested by an app. This may have also led to more in-depth privacy considerations among users.

\paragraph{\texttt{place} attribute}
\begin{figure}[!htb]
    \centering
    \includegraphics[width=0.48\textwidth]{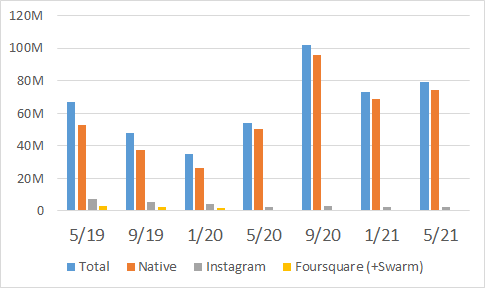}
    \caption{Sources of tweets with the \texttt{place} attribute set by month (``Total'' = all tweets with \texttt{place} collected in the 1\% sample of that month).}
    \label{fig:place_sources}
\end{figure}
We then performed the same analysis for the \texttt{place} attribute, see figure \ref{fig:place_sources}. We see the same seasonal effects here, but not the same decrease as in the previous experiment, indicating that the cause for the \texttt{coordinates} drop was in fact the policy change. The higher rate of total tweets with a \texttt{place} attribute is probably due to a higher total tweet volume starting in 2020\footnote{\url{https://blog.gdeltproject.org/visualizing-twitters-evolution-2012-2020-and-how-tweeting-is-changing-in-the-covid-19-era/}}. Fortunately, this means that the \texttt{place} attribute is still usable for research. We do see a decrease for Instagram crossposts which contain joint \texttt{coordinates} and \texttt{place}, confirming our suspicion in the previous section.

\begin{figure}[!htb!]
    \centering
    \includegraphics[width=0.48\textwidth]{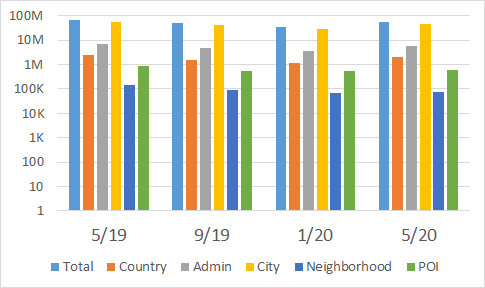}
    \caption{Share of \texttt{place\_type} in the \texttt{place} attribute by month, note log scaling (``Total'' = all tweets with \texttt{place} collected in the 1\% sample of that month).}
    \label{fig:place_granularity}
\end{figure}

Secondly, we also considered the \texttt{place\_type} field more closely to find out more about the level of detail provided by the \texttt{place} attribute. As shown in figure \ref{fig:place_granularity}, the most frequent type is ``city'' (which is also the level automatically set for Instagram crossposts). However, the most fine-grained type ``poi'' also makes up 1-2\% of the tweets, overall resulting in a still relatively large amount of available tweets with a geolocation at this granularity.

\paragraph{Locations mentioned in \texttt{text} attribute}

\begin{figure}[!htb!]
    \centering
    \includegraphics[width=0.48\textwidth]{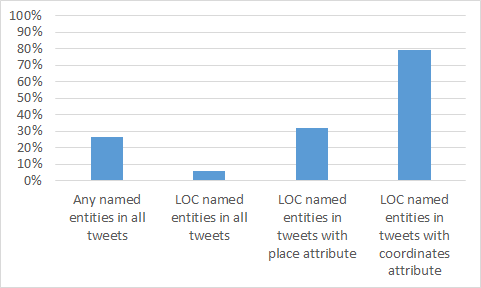}
    \caption{Shares of tweets with named entities and named LOC entities detected in one-week sample from August 2021, plus shares of tweets with LOC entities out of those with set \texttt{place} and \texttt{coordinates} attributes.}
    \label{fig:ner}
\end{figure}
Finally, we wanted to obtain a rough estimate of the amount of geolocation information contained in the \texttt{text} attribute of tweets. To this end, we performed Named Entity Recognition on the one-week sample from August 2021 described above. We used a pre-trained HuggingFace model based on RoBERTa embeddings\footnote{\url{https://huggingface.co/philschmid/distilroberta-base-ner-wikiann-conll2003-3-class}}. This model was trained on the CoNLL-2003 \cite{Tjong2003} and WikiANN data sets with the entity classes taken from WikiAnn \cite{Pan2017, Rahimi2019}. It supports 176 languages.

Figure \ref{fig:ner} shows the percentages of tweets that contain any named entity (around 26\%) and those with at least one detected LOC entity (around 6\%). We also show these numbers separately for geolocated tweets: Out of tweets with the \texttt{place} attribute, around 32\% contained LOC entities, while 79\% of tweets with the \texttt{coordinates} attribute did. This indicates a stronger semantic focus on the user's location when setting these attributes, but it also means that a large number of tweets mentioning a location has so far not been exploited. While 6\% does not sound like a lot, this set of tweets is still around 30 times as large as the number of all geolocated tweets in this sample (compare figure \ref{fig:geoloc_availability}), and that is only for direct recognition of known geographic entities. We believe this number would be even higher if other clues about location in the text were included.

\section{Effect on research}\label{sec:effect}







Previous research using geolocated tweets has mainly exploited the \texttt{coordinates} attribute under the assumption that it would contain the physical location of the device the tweet was sent from. This approach is easy to motivate - the more finegrained the location, the more information can potentially be gathered from it, even if the task at hand could also be solved with a coarser location. However, even before 2019, there were two disadvantages to this approach. On the one hand, the \texttt{coordinates} attribute is only filled in .01-.05\% of the tweets (as opposed to .5-2\% for the \texttt{place} attribute). On the other hand, crossposts from other sources filled the \texttt{coordinates} attribute differently. Most prominently, the 5-15\% of geolocated tweets coming from Instagram never contained the GPS location of the user.

Since 2019, the \texttt{coordinates} attribute cannot reliably be used to determine location anymore, as we will detail in the next section. In this section, we will discuss what this means for some common research tasks. As \cite{Middleton2018} describe, typical stakeholders of Twitter analysis include journalists, civil protection agencies or governing bodies, and businesses. We would add scientists from other domains to this list, leading to the following use cases that have been in the focus of research:
\begin{description}
\item[POIs]
POIs have been in the focus of research for the purpose of recommender systems, detecting novel or unknown POIs, analyzing opinions and possible improvements etc. Despite the availability of POI-level geolocations in the \texttt{place} attribute, researchers have mostly used the \texttt{coordinates} attribute for this purpose, presumably to gather a wider range of POIs that are not dependent on catalogs of geolocation providers \cite{Hu2013, Maeda2016}. With the loss of exact GPS coordinates, these approaches cannot easily be applied to current data. However, they could easily be adapted to POIs provided in the \texttt{place} attribute for many use cases. The exception to this are the discovery of new POIs as well as the analysis of user behavior in the vicinity of POIs \cite{Hamstead2018, Lloyd2017}. 
\item[Mobility]
Another strong focus of social media research is the analysis of human mobility, e.g. travel or commuting patterns. As before, the \texttt{coordinates} attribute has mainly served this purpose to allow for a flexible detection of origins and destinations \cite{GrantMuller2015}. Future strategies without this attribute depend on the scale of mobility to be analyzed. When tracking movement between cities or even international travel, the city-level locations in the \texttt{place} attribute should suffice. Analysis on the sub-city level is more difficult now. For a general idea, e.g. for transport optimization, POI-level locations could be exploited if a sufficient number of them is available and well-distributed across the area \cite{Huang2016}. In Instagram crossposts, a location mapped to the \texttt{coordinates} attribute can also be a street, so this may be a valid source to determine travel in the city (after excluding centroids of other places, e.g. cities). 

\item[Disasters]
Natural and man-made disasters are among the most strongly researched applications of geolocated social media. Tasks include the automatic detection of events, detection of tweets related to disasters, classification of such tweets into certain categories, and detection of actionable tweets \cite{Kruspe2021}. As before, most existing approaches use the \texttt{coordinates} attribute. This is particularly critical for use cases where action is necessary, e.g. calls for help, or where localized developments of a disaster are detected. Due to the low share of tweets with exact coordinates even before 2019, efforts have been made to determine such locations from other sources, e.g. \cite{Singh2019}. On the more general level necessary for detecting events and disaster-related tweets in the first place, we believe city or POI locations from the \texttt{place} attribute will often be sufficient.

\item[Public health]
Similar to the disaster topic, social media has also been suggested to explore public health topics such as the spread of infectious diseases \cite{Achrekar2011, Padmanabhan2014}. For this task, insights are not commonly gained on a sub-city level. Previous publications still often exploit the \texttt{coordinates} attribute, but then map it to a city or area. This could easily be substituted with the \texttt{place} attribute. Even the user-provided location could be sufficient here as most tasks are not reliant on user location change over short time spans. There are some use cases where this might be necessary, e.g. when attempting to model COVID-19 spread on a person-by-person basis, but we would argue that not a high-enough percentage of the population uses geolocated social media to be feasible.

\item[Marketing]
Marketing tasks on social media include e.g. the analysis of sentiments towards brands or products or the prediction of sales based on user expressions. This appears to be an area where exact geolocations do not serve a purpose and should therefore not see any detriment from the change. In fact, most research so far has focused on analysis without any geo-based statistics or on the country level, e.g. \cite{Jendoubi2020, Ibrahim2019, Lassen2014}.

\item[Politics and social sciences]
In politics and social sciences, geolocations are usually not required at a very fine-grained level. The most prominent task in political social media analysis, election prediction, cannot even produce results comparable to the actual election result beyond the city/area level, which is affirmed by the overview provided in \cite{Gayo2012}. Similarly, empirical analysis of social effects or opinions usually operates on a larger scale, and requires inputs from city areas or larger, e.g. \cite{Ceron2014, Kling2012}, this may be an area where the \texttt{place} attribute may be useful at the \texttt{neighborhood} level or lower. A notable exception is \cite{Hobbs2018}, which focuses on a quantitative analysis of geolocation provision by Arabic/Muslim users over time influenced by safety concerns. Another task where location is crucial is the recognition of suicide risk in users, where recognizing their location could serve to provide help \cite{Jashinsky2013} (also related to the Public Health topic).

\item[Mapping]
The most critical application of geolocated social media research seems to be mapping. So far, this task has been almost completely reliant on exact coordinates. Naturally, if we want to detect novel geographic structures or mapping details about known ones (e.g. building usages \cite{Haeberle2019}), we require the exact location the users are talking about in tweets. Future research therefore needs to detect these locations in other ways, some of which are suggested in the next section. There are some tasks where researchers may be able to rely on known places from geolocation service (i.e. the \texttt{place} attribute), e.g. collection of usage statistics over time \cite{FriasMartinez2014}. There is also a close relation to the POI tasks described above.

\end{description}

\section{Suggestions for future research}\label{sec:suggestions}




As we saw in the previous sections, the availability of geolocations in Twitter data has changed quite drastically since 2019. One main takeaway here is that the \texttt{coordinates} attribute, if it is filled, does not signify a user's physical location when they made a post anymore. As of now, this attribute is only available in Instagram crossposts, where it is set to the centroid of pre-defined locations coming from Instagram. Moreover, this was already the case for Instagram crossposts before 2019, meaning that for around 10\% of tweets, the \texttt{coordinates} attribute never contained the user's GPS location in the first place. In the future, researchers should therefore not rely on this attribute as a source of exact geolocation anymore.

Moving forward, researchers need to carefully consider which level of granularity is necessary for the task at hand. As a general rule, the more finegrained, the less data is available. We suggest the following sources of geolocation depending on the level required:
\begin{description}
\item[Country or city level] This is the easiest level to obtain. Nearly all geolocated tweets, whether coming from native Twitter or from Instagram, currently contain location information on the city level or finer in the \texttt{place} attribute.
\item[Point of interest (POI)] There are currently two ways to obtain tweets tagged at the POI level:
\begin{enumerate}
\item Tweets coming from native Twitter can directly contain a POI location in the \texttt{place} attribute, including the POI's name and bounding box. This is the case for around 1-2\% of all geolocated tweets.
\item Tweets coming from Instagram contain a centroid in the \texttt{coordinates} attribute that often corresponds to a POI. Unfortunately, this centroid first needs to be mapped back to a POI. In principle, this is possible using geolocation services such as those from Yelp, Foursquare, or Twitter's own reverse geocoding service\footnote{\url{https://developer.twitter.com/en/docs/twitter-api/v1/geo/places-near-location/overview}}. This process may introduce errors, and requires disentangling POI centroids from those for cities or countries, but it may be worth it to obtain more POI-level data. According to \cite{Maurer2020}, around 70\% of Instagram crossposts contain geolocations at this granularity.
\end{enumerate}
\item[More fine-grained or use case-specific] A finer level of granularity, e.g. for specific buildings or geographic structures other than POI, is not currently widely available via the geolocation data directly provided within tweets. The only way to potentially obtain this information lies in analyzing the text content of the tweet, which we will discuss further below.
\end{description}
Another aspect that researchers need to keep in mind now is that none of these locations are necessarily the physical spot where the tweet was sent, but a place that the user chose to attach to the tweet. In the case of native Twitter posts, these locations will at least be somewhere close to the GPS location of the device (around 200m radius), whereas in Instagram, they may be anywhere in the world. This can be an advantage in certain scenarios, though, allowing to take information into account even though the poster was not physically present at the location they are discussing.

In general, the percentage of geolocated tweets out of all tweets is low at 1-2\%. We believe that there is a large amount of untapped information for tasks that require a geolocation within the remaining 98-99\%. Exploiting this data would require determining the tweets' location from other sources, most prominently the actual content of the tweet. The simplest approach consists of performing Named Entity Recognition (NER) on the texts to detect known locations; in our preliminary experiments, we found that around 6\% of the texts of all tweets contained geographic entities, out of which only about 3.5\% were already covered by geolocated tweets. In a second step, these entities then need to be mapped to coordinates, the so-called geocoding, with the possible difficulty of having to disambiguate entity names. As a side note, the same process can be applied to locations set in user profiles (30-40\% of profiles) without performing the NER step. We need to keep in mind that this location may not be accurate for all tweets of this user, but for some tasks, it may even make more sense to work with user location rather than tweet location.

To cover an even higher percentage of tweets, geocoding can also be performed via an analysis of latent factors of the tweet text, e.g. local slang or mentions of non-geographic, but locatable entities such as sports teams. An interesting approach would lie in correlating tweet texts with known descriptions of places, e.g. from Yelp or Wikidata, or in detecting tweets for specific locations by anchoring them on known ones via few-shot learning \cite{Kruspe2019}. Other tweet metadata, such as the language, can also be taken into account. Geocoding of tweets has been a research topic for some years now, such as in W-NUT's own shared task in 2016 \cite{Han2016} (for other examples, see e.g. \cite{Schlosser2021, Paule2019}). Image content that is now part of many tweets, especially Instagram crossposts, could also be analyzed with computer vision models as a source of location.

When using geocoding approaches, we cannot be sure what level of granularity to expect, but there may be tasks where it even makes sense to leave this distinction up to the users themselves. More importantly, careful consideration is necessary here to ensure that determining the geolocation does not infringe upon the users' privacy when they have not explicitly provided this location themselves. 

\section{Conclusion}\label{sec:conclusion}
In this paper, we gave a detailed overview over the effects of Twitter's geolocation policy change in 2019. We first described the roles of the various tweet attributes provided by Twitter's API and how they are filled by Twitter itself as well as third-party apps, in particular Instagram. We point out a particular issue with the assumption that the \texttt{coordinates} attribute contains the exact location of the user, which has never been the case for Instagram crossposts.

Next, we calculated a range of statistics, including a verification that the amount of tweets with GPS locations has starkly decreased since the policy change. We also showed that the \texttt{place} attribute is still usable and more broadly available, albeit less fine-grained, and that the text content of tweets also provides a lot of useful clues to determine geolocation. Future research could elucidate the usage of various types of (explicit or implicit) geotagging depending on user demographics.

We then discussed the effect on different research tasks and conclude that there are many cases where GPS granularity is not necessary, which is also important because of ethical data minimization principles. Exceptions include mapping, tasks where users require immediate help in-person, and certain mobility analyses.

Finally, we suggest what technical steps could be taken moving forward, depending on the required level of geolocation granularity. Besides the explicit availability of locations, geocoding approaches based on tweet content are a promising research direction that could unlock a large percentage of the Twitter stream for geo-based tasks.



\section*{Acknowledgements}
 This work is supported by the European Research Council (ERC) under the European Union's Horizon 2020 research and innovation programme (grant agreement No. [ERC-2016-StG-714087], Acronym: \textit{So2Sat}).
 
 We would like to thank Auxane Boch for ethical insights into the issue.

\bibliography{anthology,references}
\bibliographystyle{acl_natbib}



\end{document}